\documentclass[apj]{emulateapj}

\shorttitle{Active Region Outflows and the Solar Wind}
\shortauthors{Brooks}
\begin{document}

\title{Establishing a Connection Between Active Region Outflows and the Solar Wind: 
Abundance Measurements with EIS/{\it Hinode}}
\author{David H. Brooks \altaffilmark{1}} 
\affil{College of Science, George Mason University, 4400 University Drive, Fairfax, VA 22030}                                  
\email{dhbrooks@ssd5.nrl.navy.mil}
\author{Harry P. Warren}
\affil{Space Science Division, Naval Research Laboratory, Washington, DC 20375}
\altaffiltext{1}{Present address: Hinode Team, ISAS/JAXA, 3-1-1 Yoshinodai, Sagamihara, Kanagawa 229-8510, Japan}

\begin{abstract}
One of the most interesting discoveries from 
{\it Hinode} is the 
presence of persistent high temperature high speed outflows from the edges of active regions. EIS measurements 
indicate that the outflows reach velocities of 50\,km~s$^{-1}$ with spectral line asymmetries approaching 200\,km~s$^{-1}$. It has been 
suggested that these outflows may lie on open 
field lines that connect to the heliosphere, and that they could potentially be a significant source of the slow speed 
solar wind. A direct link has been difficult to establish, however. We use EIS measurements
of spectral line intensities 
that are sensitive to changes in the relative abundance of \ion{Si}{0} and \ion{S}{0} as a result of the first
ionization potential (FIP) effect,
to measure the chemical composition in the outflow regions of AR 10978 over a 5 day period
in December 2007. We find that \ion{Si}{0} is always enhanced over \ion{S}{0} by a factor
of 3--4. This is generally consistent with the enhancement factor of low FIP elements measured {\it in-situ} in the slow
solar wind by non-spectroscopic methods. Plasma with a slow wind-like composition was therefore flowing from the edge
of the active region for at least 5 days. Furthermore, on December 10--11, when the outflow from the western side 
was favorably oriented in the Earth direction, the \ion{Si}{0}/\ion{S}{0} ratio was found to match the value measured
a few days later by ACE/SWICS. These results provide strong observational evidence for 
a direct connection between the solar wind, and the coronal plasma in 
the outflow regions.
\end{abstract}
\keywords{Sun: corona---Sun: abundances--- solar wind}

\section{Introduction}
Recent observations by the EUV imaging spectrometer \citep[EIS,][]{culhane_etal2007a} 
and X-ray telescope \citep[XRT,][]{golub_etal2007} on {\it Hinode} \citep{kosugi_etal2007}
have detected the presence of high temperature outflows at the edges of active
regions (\citealt{sakao_etal2007,doschek_etal2007,harra_etal2007}). These outflows show 
velocities on the order of tens of km s$^{-1}$ and the high spectral resolution EIS data have revealed 
that the line profiles have asymmetries that reach
several hundred km s$^{-1}$ \citep{bryans_etal2010,peter_2010}. 
Similar outflows 
have been observed previously at lower temperatures as intensity perturbations in TRACE images 
\citep{schrijver_etal1999}, or Doppler shifts in SUMER spectra 
\citep{marsch_etal2004}, but the relationship between the cool and hot flows is only now being established
through comprehensive studies using the broad temperature coverage of EIS \citep{delzanna_2008,warren_etal2010b}. 
Active regions are thought to be possible sources of the slow solar wind, especially during solar maximum,
and there have
been several studies pursuing this connection 
\citep{neugebauer_etal2002,schrijver&derosa_2003,liewer_etal2004}. It is important, however, to determine exactly
where the slow speed wind is originating from in an active region. \citet{wang_etal2009} have stressed that 
the conditions at the base of open field lines greatly influence the properties of the wind in
hydrodynamic models. The
persistence of the outflows for several days has led to the suggestion that they specifically could be the most significant
contributors
\citep{sakao_etal2007,harra_etal2008,doschek_etal2008}. A direct link has been
difficult to establish, however, and further studies are needed,
together with observational and theoretical investigations of the origin and driver 
of the outflows \citep{baker_etal2009,murray_etal2010}.

One capability of EIS that has not yet been fully exploited and could help in establishing a connection, is the ability
to measure the chemical composition of the outflows. It is known from {\it in-situ} measurements of the ion composition in
the slow speed solar wind that
elements with a first ionization potential (FIP) below about 10 eV are enhanced 
by factors of 3--4 relative to their photospheric abundances \citep{vonsteiger_etal2000,feldman&widing_2003}. 
In contrast, the fast speed solar wind shows only a small enhancement; perhaps a factor of 1.5 \citep{vonsteiger_etal2000}, and
this is also consistent
with spectroscopic measurements in the coronal hole source regions that show 
abundances that are close to photospheric \citep{feldman&laming_2000}. 
For a discussion of explanations for the FIP effect see, e.g., \citet{laming_2004}.
The magnitude of the FIP effect also appears
to be related to the coronal plasma confinement time \citep{feldman&widing_2003}, so
the source of the plasma that flows to the slow
wind must be confined long enough to reach an enhancement factor of 3--4 and then must be released to the solar wind
along open field lines. 

\begin{figure*}
\centering
\includegraphics[angle=90,width=0.99\linewidth]{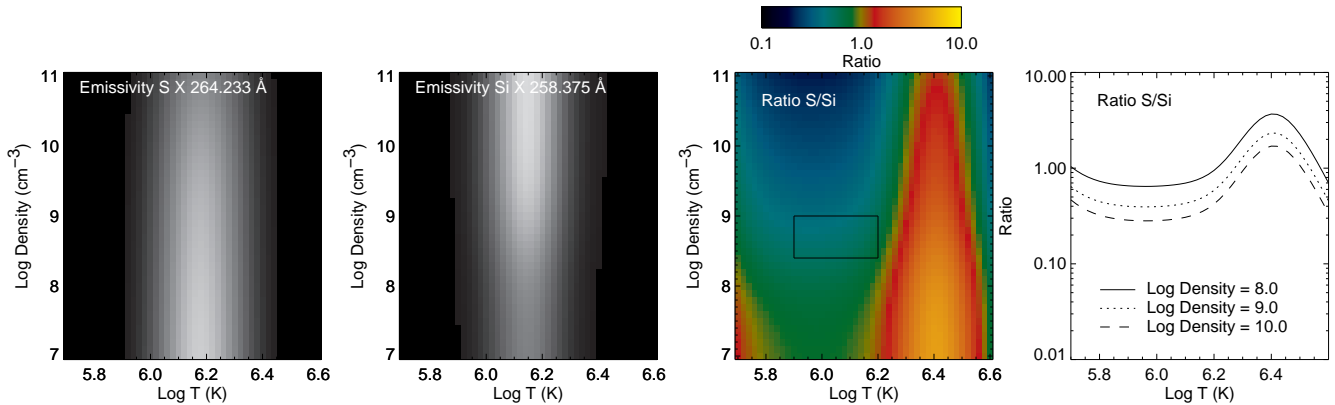}
\caption{\ion{Si}{10} 258.374\,\AA/\ion{S}{10} 264.233\,\AA\, abundance diagnostic ratio.
From left to right: emissivity as a function of temperature and density for \ion{S}{10} 264.233\,\AA\, 
and \ion{Si}{10} 258.374\,\AA. Ratio as a function of temperature and
density and as a function of temperature for 
densities of $\log$ (N$_e$/cm$^{-1}$) = 8--10. The boxed area indicates the [T$_e$,N$_e$] space derived
for the majority of the outflows. 
\label{fig1}}
\end{figure*}
\begin{figure}[ht]
\centering
\includegraphics[width=0.65\linewidth,viewport= 50 70 210 310,clip]{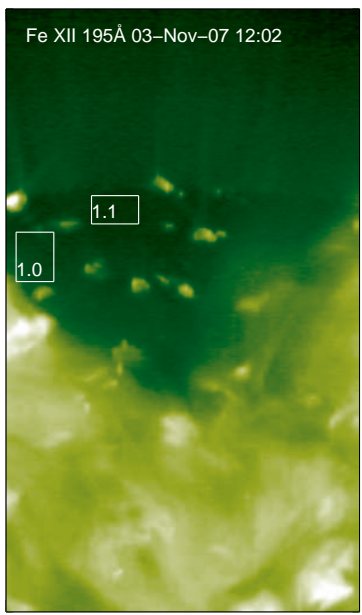}
\caption{North polar coronal hole observation used as a test of the method for deriving the FIP bias. 
The derived values within the boxes are shown.
\label{fig2}}
\end{figure}
Since the 
magnitude of the FIP effect varies substantially between different solar features 
\citep{feldman&widing_1993,sheeley_1995,sheeley_1996,
raymond_etal1997},
the possibility exists that measurements of the magnitude of the FIP effect could identify the source location of the slow wind.
Some previous work measuring the FIP-bias (ratio of coronal abundance of a low FIP element to that of a high FIP element)
in the boundary between an active region and a coronal hole has been undertaken by \citet{ko_etal2006}, who found that this
could be a possible source location of the solar wind. Similar studies are rare, however, and
to date there have been no measurements of relative abundances in the high speed
outflows near active regions. Recently \citet{feldman_etal2009b}
outlined how EIS observations of \ion{Si}{0} and \ion{S}{0} lines could be used to
measure the FIP bias at temperatures near 1.5\,MK, the peak temperature for
the active region outflows. They did not study the outflows, but gave a few illustrative
calculations for a number of targets. They also noted that accurate measurements would require
differential emission measure (DEM) analysis. In this letter, we present the
methodology needed to account for the temperature and density sensitivity of
the emission lines involved and calibrate it with measurements in several
polar coronal holes. We then measure the FIP-bias in the outflow regions of AR
10978 over a period of 5 days in December, 2007 and show that the results
are consistent with the {\it in-situ} measurements.

\section{Data processing and methodology}
The EIS instrument observes in two wavelength bands: 171--212\,\AA\, and 245--291\,\AA. It 
has 1$''$ spatial pixels and a spectral resolution of 22.3m\,\AA. The instrument is described in detail
by \citet{culhane_etal2007a}. In this letter, we analyze observations of AR 10978 obtained between December 10th and 
15th, 2007. This region has previously been studied in detail by several authors \citep{doschek_etal2008,brooks_etal2008,warren_etal2008a,ugarteurra_etal2009,bryans_etal2010}. 
The data we use were obtained with the 1$''$ slit in scanning mode. The observing sequence covers a large FOV of 
460$'' \times$384$''$
with 40s exposures at each position and we use data from 5 runs of this sequence. 
Calibration and processing of the data were performed using standard procedures in SolarSoft. In addition, the orbital drift 
of the spectrum on the detector due to instrument thermal variations and spacecraft revolution were
corrected using the artificial neural network model of \citet{kamio_etal2010}. 
This model also corrects the spatial offsets between detectors and the spectral curvature
caused by the grating tilt. 
It uses instrument temperature 
information and spacecraft housekeeping data to perform the correction, and the residual uncertainty of the wavelength positions
is expected to be $\sim$4.5\,km~s$^{-1}$. The reference wavelength is taken from an average of all the mission data
for the \ion{Fe}{12} 195.119\,\AA\, line, but we make an additional correction for \ion{Fe}{13} 202.044\,\AA\, using
the average value obtained in a relatively quiet area of the rasters. We use the lowest 50 pixels for this purpose.

\begin{figure*}
\centering
\includegraphics[width=1.00\linewidth]{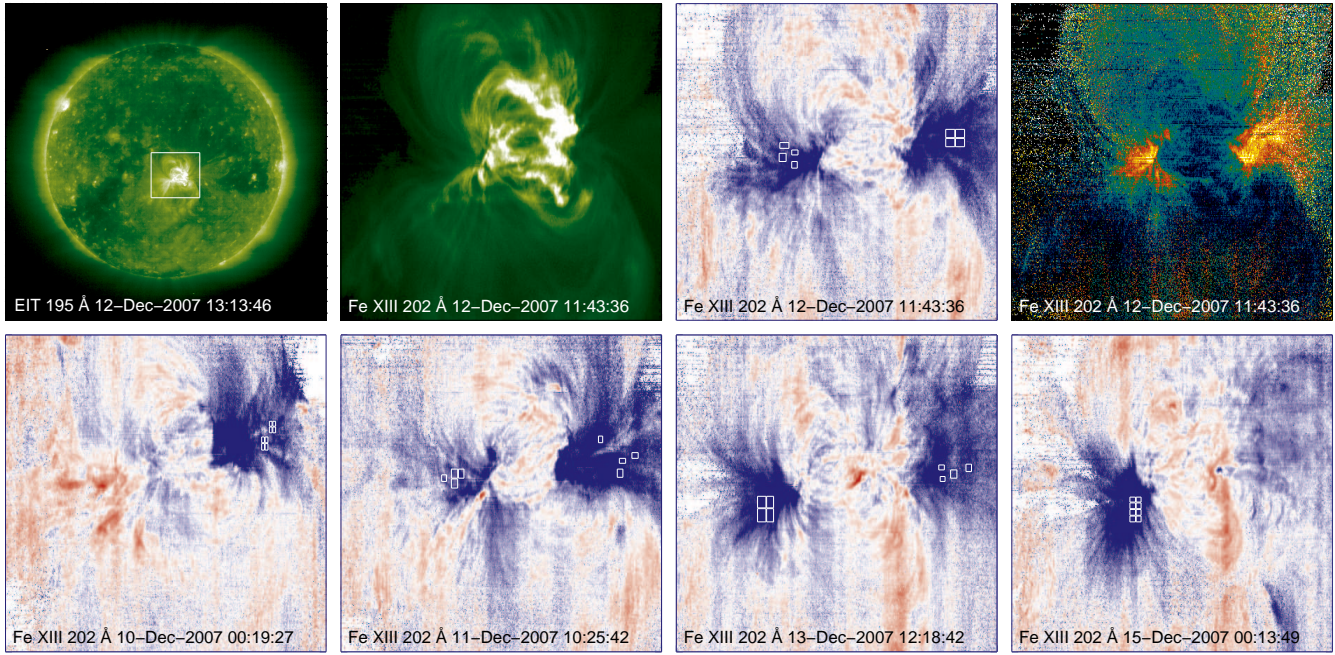}
\caption{
Context images and velocity maps for AR 10978. Top row: examples for December 12: 
EIT 195\,\AA\, filter image, EIS \ion{Fe}{13} 202.044\,\AA\, intensity image,
\ion{Fe}{13} 202.044\,\AA\, Doppler velocity map, and 
\ion{Fe}{13} 202.044\,\AA\, non-thermal velocity map. 
Bottom row: velocity maps for the other 4 days between December 10 and 15. The areas used for the
analysis are shown by small boxes on the velocity maps.
\label{fig3}}
\end{figure*}
The high spectral resolution of EIS 
enables observation of coronal emission line profiles in detail. 
In previous work we have found the line widths in the core of an active region to be narrow \citep{brooks&warren_2009}, however,
several studies have identified
blue-wing asymmetries associated with different solar features including the outflows \citep{hara_etal2008,depontieu_etal2009,bryans_etal2010,peter_2010}. 
A Gaussian function tends to broaden and 
shift towards the wings
to account for this asymmetry, but
here we mainly use the derived velocities to identify the outflow regions and are more concerned with
the accuracy of the intensity measurements. 
Since the contribution of the asymmetry to the total line intensity is generally small,
we fit the spectral features with single and multiple Gaussians.

\begin{deluxetable*}{lccccccc}
\tabletypesize{\scriptsize}
\tablewidth{0pt}
\tablecaption{Properties of AR 10978 outflows measured in EIS slit scans.}
\tablehead{
\multicolumn{1}{l}{Date} &
\multicolumn{1}{c}{Start Time} &
\multicolumn{1}{c}{Location} &
\multicolumn{1}{c}{$v$/km s$^{-1}$} &
\multicolumn{1}{c}{$\eta$/km s$^{-1}$}  &
\multicolumn{1}{c}{$\log$ (N$_e$/cm$^{-3}$)}  &
\multicolumn{1}{c}{$\log$ (T$_p$/K)}  &
\multicolumn{1}{c}{f$_{FIP}$} 
}
\tablenotetext{}{$v$ - Doppler velocity.}
\tablenotetext{}{$\eta$ - non-thermal velocity.}
\tablenotetext{}{$N_e$ - electron density.}
\tablenotetext{}{$T_p$ - temperature of emission measure peak.}
\tablenotetext{}{f$_{FIP}$ - FIP bias.}
\startdata
10-Dec-2007 & 00:19:27 & West & -18.3 &  41.9 &  8.6 & 6.1 & 3.0 \\
            &          &      & -23.9 &  51.6 &  8.5 & 6.1 & 2.8 \\
            &          &      & -19.1 &  42.0 &  8.5 & 6.0 & 2.7 \\
            &          &      & -20.4 &  46.9 &  8.5 & 6.2 & 3.0 \\
            &          &      & -17.5 &  35.9 &  8.4 & 6.2 & 2.9 \\
            &          &      & -16.8 &  38.6 &  8.5 & 6.2 & 2.7 \\
            &          &      & -17.4 &  37.1 &  8.5 & 6.2 & 2.8 \\
            &          &      & -17.8 &  39.4 &  8.5 & 6.2 & 2.5 \\
11-Dec-2007 & 10:25:42 & East & -10.9 &  33.0 &  8.8 & 6.3 & 3.7 \\
            &          &      & -10.9 &  34.8 &  8.8 & 6.2 & 3.8 \\
            &          &      &  -9.8 &  31.7 &  8.7 & 6.3 & 3.5 \\
            &          &      & -14.2 &  38.1 &  8.8 & 6.2 & 3.9 \\
            &          & West & -26.8 &  50.7 &  8.5 & 6.2 & 3.7 \\
            &          &      & -17.9 &  39.2 &  8.4 & 6.2 & 3.3 \\
            &          &      & -24.1 &  45.6 &  8.6 & 6.2 & 3.2 \\
            &          &      & -21.4 &  45.8 &  8.5 & 6.2 & 3.3 \\
12-Dec-2007 & 11:43:36 & East & -16.6 &  39.7 &  8.7 & 6.2 & 4.0 \\
            &          &      & -12.6 &  32.9 &  8.6 & 6.2 & 3.5 \\
            &          &      & -17.3 &  39.0 &  8.6 & 5.6 & 3.8 \\
            &          &      & -20.4 &  41.0 &  8.7 & 6.3 & 4.1 \\
            &          & West & -18.1 &  40.4 &  8.5 & 6.2 & 3.1 \\
            &          &      & -20.8 &  43.3 &  8.5 & 6.2 & 3.7 \\
            &          &      & -21.8 &  45.8 &  8.5 & 6.2 & 3.4 \\
            &          &      & -22.3 &  47.3 &  8.5 & 6.2 & 3.8 \\
13-Dec-2007 & 12:18:42 & East & -20.7 &  35.5 &  8.7 & 6.2 & 3.6 \\
            &          &      & -21.1 &  41.9 &  8.8 & 5.9 & 3.6 \\
            &          &      & -21.2 &  35.9 &  8.7 & 6.2 & 3.4 \\
            &          &      & -26.4 &  47.8 &  8.7 & 5.6 & 3.9 \\
            &          & West & -17.2 &  37.0 &  8.4 & 6.2 & 3.5 \\
            &          &      & -12.5 &  35.6 &  8.4 & 6.2 & 2.9 \\
            &          &      & -15.3 &  35.0 &  8.4 & 6.2 & 2.7 \\
            &          &      & -21.9 &  43.4 &  8.5 & 6.2 & 2.8 \\
15-Dec-2007 & 00:13:49 & East & -41.2 &  57.5 &  8.7 & 6.0 & 4.0 \\
            &          &      & -27.5 &  41.5 &  8.8 & 6.2 & 3.9 \\
            &          &      & -41.7 &  57.4 &  8.8 & 6.0 & 3.7 \\
            &          &      & -33.2 &  47.2 &  8.8 & 6.2 & 4.1 \\
            &          &      & -40.0 &  59.4 &  8.7 & 5.6 & 3.9 \\
            &          &      & -35.1 &  53.1 &  9.0 & 6.0 & 3.9 \\
            &          &      & -32.7 &  47.4 &  8.7 & 6.0 & 3.8 \\
            &          &      & -24.8 &  46.3 &  8.8 & 6.0 & 3.7 \\
\enddata
\label{tab}
\end{deluxetable*}
A number of methods can be used to determine the FIP-bias ($f_{FIP}$) in the outflows and a detailed
discussion of diagnostic ratios in the EIS wavelength range is given by \citet{feldman_etal2009b}.
They show that the 
\ion{Si}{10} 258.374\,\AA/\ion{S}{10} 264.233\,\AA\, ratio is
constant to within $\sim$30\% in the $\log$ (T$_e$/K) = 6.0 to 6.2 range, which makes it useful for
analysis of the outflows. We recomputed the ratio 
using the CHIANTI v6.0.1 database
\citep{dere_etal1997,dere_etal2009}, and 
show it as a function
of temperature and density in Figure \ref{fig1}. 
With these data we find that the ratio 
varies by $\sim$40\% in the $\log$ (T$_e$/K) = 5.7 to 6.2 range, but deviates strongly
at high temperatures. In regions
with a significant high temperature emission measure then, the ratio should properly be convolved with the
DEM distribution. 
We also find a significant sensitivity of the ratio to the electron
density (factor of 2.3 between $\log$ (N$_e$/cm$^{-3}$) = 8 and 10). So the density in the target
region also needs to be measured and accounted for.
As we will show below, the densities determined for the outflows do not vary sufficiently to cause
a greater than 30\% change in the ratio. 

We adopt the following procedure for our analysis. First, we measure the density in the outflow region using the
\ion{Fe}{13} 202.044\,\AA/\ion{Fe}{13} 203.826\,\AA\, diagnostic ratio. Then, we derive the
DEM distribution using a series of \ion{Fe}{8}--\ion{}{16} lines to minimize uncertainties due to elemental
abundances. The specific lines used are:
\ion{Fe}{8} 185.213\,\AA,
\ion{Fe}{9} 188.485\,\AA, \ion{Fe}{10} 184.536\,\AA, \ion{Fe}{11} 188.216\,\AA, \ion{Fe}{12} 195.119\,\AA,
\ion{Fe}{13} 202.044\,\AA, \ion{Fe}{14} 274.203\,\AA, \ion{Fe}{15} 284.160\,\AA, and \ion{Fe}{16} 262.984\,\AA.
The DEM is reconstructed using the Markov-Chain Monte Carlo (MCMC) algorithm distributed with the
PINTofALE spectroscopy package \citep{kashyap&drake_1998,kashyap&drake_2000}. For all the calculations we
adopted the photospheric abundances of \citet{grevesse_etal2007}.
The atomic data for this calculation were computed using the CHIANTI database at the fixed electron density
previously measured for each outflow. Once the DEM is computed, the \ion{Si}{10} 258.374\,\AA\, and
\ion{S}{10} 264.233\,\AA\, line intensities are calculated. Since \ion{Si}{0} and \ion{Fe}{0} are both low FIP
elements, the calculated \ion{Si}{10} 258.473\,\AA\, intensity should be well matched, but we scale
the \ion{Fe}{0} DEM (if necessary) to make sure that it is.
We find, however, that the difference is less than 20\% for all the outflow regions we investigate. The ratio of the calculated
to observed intensity for the high FIP \ion{S}{10} 264.233\,\AA\, line is then the FIP bias, fully accounting
for the temperature and density dependence of the emissivities.

As an independent check of the method we derived $f_{FIP}$ values for eight polar coronal
hole observations. Since the polar coronal hole is the presumed source of the fast speed
solar wind, and the chemical composition there is close to photospheric, we should obtain
values close to one if the method is working correctly. Figure \ref{fig2} shows an example
for observations taken on 2007, November 3. The EIS scan used the 2$''$ slit to cover an
area of 300$''$ by 512$''$ in around 1 hour. The exposure time was 50s. The FIP bias was
derived using spatially averaged line profiles from the indicated areas,
and found to be 1.0 and 1.1, respectively. In all
eight regions $f_{FIP}$ was found to be 1.2$\pm$15\%. This agreement with expectations
gives us confidence that the method is working correctly.

\section{Results}
Figure \ref{fig3} shows context images of AR 10978 for December 12 when it was near disk center (top row). A
{\it SOHO} Extreme ultraviolet Imaging 
Telescope \citep[EIT,][]{delaboudiniere_etal1995} full Sun image is shown with the EIS raster FOV overlaid as a box. 
Note that the preceding and following coronal holes are located outside of this FOV, so
this region is a good target for examining the outflow regions in isolation from any interaction with
the coronal hole boundaries where previous measurements of $f_{FIP}$ have been made in other 
active regions \citep{ko_etal2006}. The figure also shows intensity, 
Doppler velocity, and non-thermal velocity maps all derived from Gaussian fits
to the \ion{Fe}{13} 202.044\,\AA\, line profile. The non-thermal velocity is computed by subtraction in quadrature of the
thermal and instrumental widths. The thermal width is calculated assuming the peak temperature of each outflow (T$_p$ in Table \ref{tab}). 
The on orbit instrumental width
is assumed to be 56m\,\AA\, \citep{brown_etal2008}.

One can generally associate the dark intensity
areas at the east and west side of the active region with the blue-shifted emission 
and regions of large non-thermal velocity, though careful comparison would be needed to understand if they 
are related to each other in detail. Doppler velocity maps are also shown for December 10--15.
These maps are used to select
outflow regions for further analysis, and the chosen regions are shown by the small boxes.
Since the \ion{S}{10} line is weak in the outflows, averaging over a
small area is necessary to increase the signal-to-noise ratio. We selected 40 locations in total in both the solar east and west outflow
regions over the 5 days.

In making this selection, we only chose regions of outflow
along the line-of-sight. For example, on December 10th, the solar east side of the AR presumably has
an outflow, and this rotates into view on the 11th. On the 10th, however, this region shows only
red-shifts and low non-thermal velocities, presumably due to line-of-sight effects. On this day, 
we only chose areas in the blue-shifted west side outflow. 

To ensure that our selected locations were really in the outflows, we measured the 
Doppler velocities.
These and the
non-thermal velocities are noted in Table \ref{tab}. The results indicate that this AR
shows bulk outflows of 9.8--41.7\,km~s$^{-1}$ and non-thermal mass motions of 31.6--59.4\,km~s$^{-1}$. 

We then used the \ion{Fe}{13} line ratio to measure the electron density and the results are also
shown in Table \ref{tab}. We find values in the range $\log$ (N$_e$/cm$^{-3}$) = 8.4--9.0. These are
broadly consistent with the results found by \citet{doschek_etal2008}.
The DEM distributions at fixed electron density were used to determine the peak temperatures of the 
outflows. The results are also shown in Table \ref{tab} and fall in the range $\log$ (T$_e$/K) = 5.6--6.3. Finally,
the calculated $f_{FIP}$ measurements are also shown in Table \ref{tab} and
fall in the range 2.5--4.1. This
indicates that the FIP 
enhancement factors in the outflows are in agreement with expectations from the {\it in-situ}
measurements in the slow wind.

\section{Summary}
Using data from {\it Hinode} EIS we have studied the outflow regions of AR 10978 over 
5 days in December 2007. 
We find that the outflows show Doppler velocities of -22\,km~s$^{-1}$ and mass motions
of 43\,km~s$^{-1}$ on average. We also measured the electron density and temperature in the outflows and found 
average values of $\log$ (N$_e$/cm$^{-1}$) = 8.6 and $\log$ (T$_e$/K) = 6.2, respectively. Combining
an emission measure analysis with the modeling of 
\ion{Si}{10} and \ion{S}{10} lines, 
we measured the FIP bias in the outflows of the active region away from any surrounding coronal holes. We found that 
\ion{Si}{0} is always enhanced over \ion{S}{0} by a factor of 2.5--4.1, with a mean value of 3.4. 
These results generally agree with the enhancement factors of low FIP elements measured {\it in-situ} in the slow
solar wind by non-spectroscopic methods, and the enrichment was
consistent throughout the observations. 
The fact that plasma with a similar composition to the slow speed wind
was continuously flowing out from the edge of the active region for several days lends strong support to 
the suggestion that the outflows contribute to the slow wind.
We therefore
show new evidence of a direct connection between the slow speed solar wind, 
and the coronal plasma in 
the outflow regions.

To conclusively prove this connection, however, one should directly compare the EIS 
\ion{Si}{0}/\ion{S}{0} ratio in the outflows with that measured in the slow wind three
days later (the travel time to Earth). If the plasma flowing from the AR really reaches
the slow wind at Earth, the {\it in-situ} measurements should match the spectroscopic
ones.
Daily averages of the \ion{Si}{0}/\ion{S}{0} ratio measured by the Solar Wind Ion Composition
Spectrometer \citep[SWICS,][]{gloeckler_etal1998} aboard 
the Advanced Composition Explorer \citep[ACE,][]{stone_etal1998} are available for us to make this comparison
for the disk passage of AR 10978. The EIS observations indicate that 
the western outflow was near central meridian and favorably
oriented towards Earth on December 10--11. The best dates for the SWICS comparison are therefore
December 13--14. 
We examined the SWICS \ion{Si}{0}/\ion{S}{0} measurements on these dates and found average values
of 2.3 and 3.5, respectively. From Table \ref{tab} we see that for December 10--11 the EIS 
averages are 2.8 and 3.4, respectively. 
The EIS results are thus within 20\% of
the SWICS measurements. Note that no connection could be established before or after these
dates, indicating that the influence of the outflows is only seen when they are near disk center.
Further work will be needed to determine if AR 10978 is a rare case, and 
also to quantify whether the outflow contribution to the slow wind is dominant or not.

Finally, we note that there
may be other areas of an active region which could
contribute to the solar wind and larger more systematic
studies are needed. We have made some preliminary measurements 
in several locations in
the core of the December
2007 region and find the FIP bias to be similar
to that of the outflows. It is difficult to
see how these closed field regions could contribute {\it directly} to
the solar wind since no blue-shifted Doppler signatures are seen there. They
could, however, contribute indirectly, for example, by
reconnecting with open field lines. At present the
outflows are the only regions that are known to meet the
two necessary conditions for direct contribution to the wind:
upflow and composition.

\acknowledgements
We would like to thank Yuan Kuen-Ko for very helpful discussions. This work was
performed under contract with the Naval Research Laboratory and was funded by
the NASA {\it Hinode} program.
{\it Hinode} is a Japanese mission developed and launched by ISAS/JAXA,
with NAOJ as domestic partner and NASA and STFC (UK) as international partners.
It is operated by these agencies in co-operation with ESA and NSC (Norway).


\begin{thebibliography}{}

\bibitem[\protect\citeauthoryear{{Baker} et~al.}{{Baker}
  et~al.}{2009}]{baker_etal2009}
{Baker}, D., {van Driel-Gesztelyi}, L., {Mandrini}, C.~H., {D{\'e}moulin}, P.,
  \& {Murray}, M.~J. 2009, \apj, 705, 926

\bibitem[\protect\citeauthoryear{{Brooks}, {Ugarte-Urra}, \& {Warren}}{{Brooks}
  et~al.}{2008}]{brooks_etal2008}
{Brooks}, D.~H., {Ugarte-Urra}, I.,  \& {Warren}, H.~P. 2008, \apjl, 689, L77

\bibitem[\protect\citeauthoryear{{Brooks} \& {Warren}}{{Brooks} \&
  {Warren}}{2009}]{brooks&warren_2009}
{Brooks}, D.~H.,  \& {Warren}, H.~P. 2009, \apjl, 703, L10

\bibitem[\protect\citeauthoryear{{Brown} et~al.}{{Brown}
  et~al.}{2008}]{brown_etal2008}
{Brown}, C.~M., {Feldman}, U., {Seely}, J.~F., {Korendyke}, C.~M.,  \& {Hara},
  H. 2008, \apjs, 176, 511

\bibitem[\protect\citeauthoryear{{Bryans}, {Young}, \& {Doschek}}{{Bryans}
  et~al.}{2010}]{bryans_etal2010}
{Bryans}, P., {Young}, P.~R.,  \& {Doschek}, G.~A. 2010, \apj, 715, 1012

\bibitem[\protect\citeauthoryear{{Culhane} et~al.}{{Culhane}
  et~al.}{2007}]{culhane_etal2007a}
{Culhane}, J.~L., et~al. 2007, \solphys, 243, 19

\bibitem[\protect\citeauthoryear{{De Pontieu} et~al.}{{De Pontieu}
  et~al.}{2009}]{depontieu_etal2009}
{De Pontieu}, B., {McIntosh}, S.~W., {Hansteen}, V.~H.,  \& {Schrijver}, C.~J.
  2009, \apjl, 701, L1

\bibitem[\protect\citeauthoryear{{Del Zanna}}{{Del
  Zanna}}{2008}]{delzanna_2008}
{Del Zanna}, G. 2008, \aap, 481, L49

\bibitem[\protect\citeauthoryear{{Delaboudiniere} et~al.}{{Delaboudiniere}
  et~al.}{1995}]{delaboudiniere_etal1995}
{Delaboudiniere}, J.-P., et~al. 1995, \solphys, 162, 291

\bibitem[\protect\citeauthoryear{{Dere} et~al.}{{Dere}
  et~al.}{1997}]{dere_etal1997}
{Dere}, K.~P., {Landi}, E., {Mason}, H.~E., {Monsignori Fossi}, B.~C.,  \&
  {Young}, P.~R. 1997, \aaps, 125, 149

\bibitem[\protect\citeauthoryear{{Dere} et~al.}{{Dere}
  et~al.}{2009}]{dere_etal2009}
{Dere}, K.~P., {Landi}, E., {Young}, P.~R., {Del Zanna}, G., {Landini}, M.,  \&
  {Mason}, H.~E. 2009, \aap, 498, 915

\bibitem[\protect\citeauthoryear{{Doschek} et~al.}{{Doschek}
  et~al.}{2007}]{doschek_etal2007}
{Doschek}, G.~A., et~al. 2007, \apjl, 667, L109

\bibitem[\protect\citeauthoryear{{Doschek} et~al.}{{Doschek}
  et~al.}{2008}]{doschek_etal2008}
{Doschek}, G.~A., {Warren}, H.~P., {Mariska}, J.~T., {Muglach}, K., {Culhane},
  J.~L., {Hara}, H.,  \& {Watanabe}, T. 2008, \apj, 686, 1362

\bibitem[\protect\citeauthoryear{{Feldman} \& {Laming}}{{Feldman} \&
  {Laming}}{2000}]{feldman&laming_2000}
{Feldman}, U.,  \& {Laming}, J.~M. 2000, \physscr, 61, 222

\bibitem[\protect\citeauthoryear{{Feldman} et~al.}{{Feldman}
  et~al.}{2009}]{feldman_etal2009b}
{Feldman}, U., {Warren}, H.~P., {Brown}, C.~M.,  \& {Doschek}, G.~A. 2009,
  \apj, 695, 36

\bibitem[\protect\citeauthoryear{{Feldman} \& {Widing}}{{Feldman} \&
  {Widing}}{1993}]{feldman&widing_1993}
{Feldman}, U.,  \& {Widing}, K.~G. 1993, \apj, 414, 381

\bibitem[\protect\citeauthoryear{{Feldman} \& {Widing}}{{Feldman} \&
  {Widing}}{2003}]{feldman&widing_2003}
{Feldman}, U.,  \& {Widing}, K.~G. 2003, Space Science Reviews, 107, 665

\bibitem[\protect\citeauthoryear{{Gloeckler} et~al.}{{Gloeckler}
  et~al.}{1998}]{gloeckler_etal1998}
{Gloeckler}, G., et~al. 1998, \ssr, 86, 497

\bibitem[\protect\citeauthoryear{{Golub} et~al.}{{Golub}
  et~al.}{2007}]{golub_etal2007}
{Golub}, L., et~al. 2007, \solphys, 243, 63

\bibitem[\protect\citeauthoryear{{Grevesse}, {Asplund}, \& {Sauval}}{{Grevesse}
  et~al.}{2007}]{grevesse_etal2007}
{Grevesse}, N., {Asplund}, M.,  \& {Sauval}, A.~J. 2007, \ssr, 130, 105

\bibitem[\protect\citeauthoryear{{Hara} et~al.}{{Hara}
  et~al.}{2008}]{hara_etal2008}
{Hara}, H., {Watanabe}, T., {Harra}, L.~K., {Culhane}, J.~L., {Young}, P.~R.,
  {Mariska}, J.~T.,  \& {Doschek}, G.~A. 2008, \apjl, 678, L67

\bibitem[\protect\citeauthoryear{{Harra} et~al.}{{Harra}
  et~al.}{2007}]{harra_etal2007}
{Harra}, L.~K., {Hara}, H., {Imada}, S., {Young}, P.~R., {Williams}, D.~R.,
  {Sterling}, A.~C., {Korendyke}, C.,  \& {Attrill}, G.~D.~R. 2007, \pasj, 59,
  801

\bibitem[\protect\citeauthoryear{{Harra} et~al.}{{Harra}
  et~al.}{2008}]{harra_etal2008}
{Harra}, L.~K., {Sakao}, T., {Mandrini}, C.~H., {Hara}, H., {Imada}, S.,
  {Young}, P.~R., {van Driel-Gesztelyi}, L.,  \& {Baker}, D. 2008, \apjl, 676,
  L147

\bibitem[\protect\citeauthoryear{{Kamio} et~al.}{{Kamio}
  et~al.}{2010}]{kamio_etal2010}
{Kamio}, S., {Hara}, H., {Watanabe}, T., {Fredvik}, T.,  \& {Hansteen}, V.~H.
  2010, ArXiv e-prints

\bibitem[\protect\citeauthoryear{{Kashyap} \& {Drake}}{{Kashyap} \&
  {Drake}}{1998}]{kashyap&drake_1998}
{Kashyap}, V.,  \& {Drake}, J.~J. 1998, \apj, 503, 450

\bibitem[\protect\citeauthoryear{{Kashyap} \& {Drake}}{{Kashyap} \&
  {Drake}}{2000}]{kashyap&drake_2000}
{Kashyap}, V.,  \& {Drake}, J.~J. 2000, Bulletin of the Astronomical Society of
  India, 28, 475

\bibitem[\protect\citeauthoryear{{Ko} et~al.}{{Ko} et~al.}{2006}]{ko_etal2006}
{Ko}, Y., {Raymond}, J.~C., {Zurbuchen}, T.~H., {Riley}, P., {Raines}, J.~M.,
  \& {Strachan}, L. 2006, \apj, 646, 1275

\bibitem[\protect\citeauthoryear{{Kosugi} et~al.}{{Kosugi}
  et~al.}{2007}]{kosugi_etal2007}
{Kosugi}, T., et~al. 2007, \solphys, 243, 3

\bibitem[\protect\citeauthoryear{{Laming}}{{Laming}}{2004}]{laming_2004}
{Laming}, J.~M. 2004, \apj, 614, 1063

\bibitem[\protect\citeauthoryear{{Liewer}, {Neugebauer}, \&
  {Zurbuchen}}{{Liewer} et~al.}{2004}]{liewer_etal2004}
{Liewer}, P.~C., {Neugebauer}, M.,  \& {Zurbuchen}, T. 2004, \solphys, 223, 209

\bibitem[\protect\citeauthoryear{{Marsch}, {Wiegelmann}, \& {Xia}}{{Marsch}
  et~al.}{2004}]{marsch_etal2004}
{Marsch}, E., {Wiegelmann}, T.,  \& {Xia}, L.~D. 2004, \aap, 428, 629

\bibitem[\protect\citeauthoryear{{Murray} et~al.}{{Murray}
  et~al.}{2010}]{murray_etal2010}
{Murray}, M.~J., {Baker}, D., {van Driel-Gesztelyi}, L.,  \& {Sun}, J. 2010,
  \solphys, 261, 253

\bibitem[\protect\citeauthoryear{{Neugebauer} et~al.}{{Neugebauer}
  et~al.}{2002}]{neugebauer_etal2002}
{Neugebauer}, M., {Liewer}, P.~C., {Smith}, E.~J., {Skoug}, R.~M.,  \&
  {Zurbuchen}, T.~H. 2002, Journal of Geophysical Research (Space Physics),
  107, 1488

\bibitem[\protect\citeauthoryear{{Peter}}{{Peter}}{2010}]{peter_2010}
{Peter}, H. 2010, \aap, 521, A51

\bibitem[\protect\citeauthoryear{{Raymond} et~al.}{{Raymond}
  et~al.}{1997}]{raymond_etal1997}
{Raymond}, J.~C., et~al. 1997, \solphys, 175, 645

\bibitem[\protect\citeauthoryear{{Sakao} et~al.}{{Sakao}
  et~al.}{2007}]{sakao_etal2007}
{Sakao}, T., et~al. 2007, Science, 318, 1585

\bibitem[\protect\citeauthoryear{{Schrijver} \& {De Rosa}}{{Schrijver} \& {De
  Rosa}}{2003}]{schrijver&derosa_2003}
{Schrijver}, C.~J.,  \& {De Rosa}, M.~L. 2003, \solphys, 212, 165

\bibitem[\protect\citeauthoryear{{Schrijver} et~al.}{{Schrijver}
  et~al.}{1999}]{schrijver_etal1999}
{Schrijver}, C.~J., et~al. 1999, \solphys, 187, 261

\bibitem[\protect\citeauthoryear{{Sheeley}}{{Sheeley}}{1995}]{sheeley_1995}
{Sheeley}, N.~R., Jr. 1995, \apj, 440, 884

\bibitem[\protect\citeauthoryear{{Sheeley}}{{Sheeley}}{1996}]{sheeley_1996}
{Sheeley}, N.~R., Jr. 1996, \apj, 469, 423

\bibitem[\protect\citeauthoryear{{Stone} et~al.}{{Stone}
  et~al.}{1998}]{stone_etal1998}
{Stone}, E.~C., {Frandsen}, A.~M., {Mewaldt}, R.~A., {Christian}, E.~R.,
  {Margolies}, D., {Ormes}, J.~F.,  \& {Snow}, F. 1998, \ssr, 86, 1

\bibitem[\protect\citeauthoryear{{Ugarte-Urra}, {Warren}, \&
  {Brooks}}{{Ugarte-Urra} et~al.}{2009}]{ugarteurra_etal2009}
{Ugarte-Urra}, I., {Warren}, H.~P.,  \& {Brooks}, D.~H. 2009, \apj, 695, 642

\bibitem[\protect\citeauthoryear{{von Steiger} et~al.}{{von Steiger}
  et~al.}{2000}]{vonsteiger_etal2000}
{von Steiger}, R., et~al. 2000, \jgr, 105, 27217

\bibitem[\protect\citeauthoryear{{Wang}, {Ko}, \& {Grappin}}{{Wang}
  et~al.}{2009}]{wang_etal2009}
{Wang}, Y., {Ko}, Y.,  \& {Grappin}, R. 2009, \apj, 691, 760

\bibitem[\protect\citeauthoryear{{Warren} et~al.}{{Warren}
  et~al.}{2008}]{warren_etal2008a}
{Warren}, H.~P., {Ugarte-Urra}, I., {Doschek}, G.~A., {Brooks}, D.~H.,  \&
  {Williams}, D.~R. 2008, \apjl, 686, L131

\bibitem[\protect\citeauthoryear{{Warren} et~al.}{{Warren}
  et~al.}{2010}]{warren_etal2010b}
{Warren}, H.~P., {Ugarte-Urra}, I., {Young}, P.~R.,  \& {Stenborg}, G.

\end{thebibliography}
\end{document}